\documentclass[preprint]{aastex}

\usepackage{color}
\usepackage{rotating}
\usepackage{lscape}

\shorttitle{Fe-K Line Time variability and Ni abundance of AGNs}
\shortauthors{Y. Fukazawa et al.}

\begin{document}

\title{Fe-K Line Time variability and Ni abundance of Distant Reflector in Seyfert Galaxies}

\author{Yasushi Fukazawa\altaffilmark{1,2,3}, 
Shun'ya Furui\altaffilmark{1}, 
Kazuma Hayashi\altaffilmark{1}, 
Masanori Ohno\altaffilmark{1,3}, 
Kazuyoshi Hiragi\altaffilmark{1}, 
and Hirofumi Noda\altaffilmark{4}}

\email{\texttt{fukazawa@hep01.hepl.hiroshima-u.ac.jp}}

\altaffiltext{1}{Department of Physical Science, Hiroshima University, 1-3-1 Kagamiyama, Higashi-Hiroshima, Hiroshima 739-8526, Japan}
\altaffiltext{2}{Hiroshima Astrophysical Science Center, Hiroshima University, 1-3-1 Kagamiyama, Higashi-Hiroshima, Hiroshima 739-8526, Japan}
\altaffiltext{3}{Core Research for Energetic Universe (Core-U), Hiroshima University, 1-3-1 Kagamiyama, Higashi-Hiroshima, Hiroshima 739-8526, Japan}
\altaffiltext{4}{Nishina Center, Institute of Physical and Chemical
Research (RIKEN), Wako, Saitama 351-0198, Japan}

\begin{abstract}

We have performed systematic studies of narrow Fe-K line (6.4 keV)
flux variability and Ni-K line intensity for Seyfert galaxies, 
using {\em Suzaku} and {\em XMM-Newton} archival data.
Significant Fe-K line variability of several tens percents was detected 
for a pair of observations separated by 1000--2000 days 
(Cen A, IC 4329 A, NGC 3516, and NGC 4151) and
 158 days (NGC 3516).
These timescales are larger by a factor of 10--100 than the inner
 radius of torus,  consistent with the view that  X-ray
 reflection by torus is a main origin of a narrow Fe-K line.
Ni-K line was detected with $>2\sigma$ level for the Circinus galaxy, 
Cen A, MRK 3, NGC 4388, and NGC 4151.
A mean and variance of 
Ni-K$\alpha$ to Fe-K$\alpha$ line intensity ratio are 0.066 and 0.026, 
respectively.
Comparing this with the Monte-Carlo simulation of reflection, Ni to Fe
 abundance ratio is 1.9$\pm$0.8 solar.
We discuss the results and possibility of Ni abundance enhancement.

\end{abstract}


\section{Introduction}

X-ray spectra of active galactic nucleus (AGNs) exhibit various
reprocessed features such as continuum absorption, fluorescence lines,
reflection humps, absorption lines, and so on.
Among them, fluorescence lines and reflection hump are prominent in the
hard X-ray band.
Reflection is believed to be mainly attributed to an accretion disk or a
distant torus \citep{Poun90,Awak91}, but a broad line region (BLR) is also suggested to be an 
origin \citep{Yaqo04,Nand06}.
Here we do not deal with a relativistic reflection from the inner
accretion disk.
The origin of torus and the connection to the BLR and
accretion disk are important topics to understand the formation of
accretion disk and the evolution of central massive black holes.

The structure of torus is not yet completely established.
Optical polarization and far infrared (FIR) observations indicate that a torus
exists with a size of $\sim$100 pc \citep{Anto93,Pier93}.
Recent near infrared (NIR) observations measured an inner torus radius as
$0.03L_{43}^{0.5}$ pc \citep{Suga06}, where $L_{43}$ is a 
luminosity of the central engine in unit of $10^{43}$ erg s$^{-1}$.
A clumpy torus is suggested to understand a large geometrical thickness 
\citep{Wils94}, FIR spectral features \citep{Nenk02},
variable X-ray absorption \citep{Risa02}, and so on.

Origin of narrow Fe-K lines in the AGN spectra is considered to be the
torus \citep{Awak91}, and a line center energy supports this view 
\citep{Yaqo04,Fuka11a}.
On the other hand, it is suggested that there is a contribution of inner
region to the narrow Fe-K lines, such as the BLR or the outer accretion
disk, 
based on a wide range of narrow Fe-K line widths \citep{Nand06,Bian08} or a
lack of Compton-thick reflection hump \citep{Ursi15}.

The variability of Fe-K line flux is also a good probe to constrain the
emission region.
Searching for Fe-K line flux variability has been attempted, but little
evidence has been reported.
This is due to that a signal-to-noise ratio of Fe-K lines is not
sufficient for the past observations, and {\em Suzaku} or {\em
XMM-Newton} observations are needed.
There are several reports about the flux variability of the Fe-K line.
A short variability timescale of sub-day was reported for MRK 841 
\citep{Petr02,Long04}.
On the other hand, an year-long timescale variability was reported for the Cen
A \citep{Fuka11b}, NGC 7213 \citep{Ursi15}, and NGC 2110 \citep{Mari15}.
\citet{Shu12} systematically studied the {\em Chandra}/HETG data of 32
AGNs and concluded that the Fe-K$\alpha$ line flux lacks a 
corresponding response to the continuum variation, but the photon
statistics are limited and a timescale on study is typically much
shorter than one year.
Therefore, systematic studies of Fe-K flux variability with good photon
statistics and various timescales are needed.

Metal abundance of reflectors can be measured from fluorescence lines of
several elements.
High-quality {\em XMM-Newton} data for the first time resolve the Ni-K
line for a few AGNs.
The Ni-to-Fe K$\alpha$ line ratio was
reported to be 0.070$\pm$0.010 for the Circinus Galaxy, 
0.13$\pm$0.03 for NGC 1068 \citep{Matt04}, and 
0.067$\pm$0.034 for MRK 3 \citep{Poun06}.
{\em Suzaku} also detected fluorescence lines other than Fe-K from bright
AGNs, such as Cen A \citep{Mark07}, MRK 3 \citep{Awak08}, 
the Circinus Galaxy \citep{Yang09}.
However, systematic measurements of fluorescence lines other than Fe-K
have not yet been performed.
Metal abundance of torus is interesting in terms of connection with BLR
and accretion disk.
Metal abundance of BLR is
reported to be supersolar by optical/UV observations \citep{Warn04}.
Based on studies of broad Fe-K and Fe-L lines associated with relativistic
reflection, a very high Fe abundance of 2--7 solar has been suggested
\citep{Zogh10,Reyn12}.

In this paper, we report systematic studies of Fe-K line flux variability 
and Ni abundances, based on archival {\em Suzaku} \citep{Mits07} 
data on Seyfert galaxies with good photon statistics, thanks to its
large effective area and repeated observations.
In addition, we utilized some archival {\em XMM-Newton} data in order to
compensate the variability timescale of 0.5--1 yr.
Throughout this paper, the errors are shown as a 90\% confidence level.
The solar abundance ratio is referred to the
solar photospheric values of 
\citet{Ande89} for photoelectric absorption, reflection, and plasma
model, and the cross-section for absorption model is set to that of 
\citet{Balu92}.

\section{Observations and Data Reduction}

We analyzed the archival {\em Suzaku} data of X-ray bright AGNs.
Selection of sample AGNs is the same as \citet{Fuka11a}; they are
detected by HXD-PIN \citep{Taka07,Koku07} above 15 keV.
This ensures a high signal-to-noise ratio around the Fe-K lines.
Also, time variability of the central engine can be traced above 10
keV freely from absorption.
In addition, X-rays above 7 keV are important to produce Fe-K line photons.
As a result, we looked at 261 observational data of 173 objects.
All the data were obtained with the standard mode of XIS 
($5\times5$ or $3\times3$) and HXD.
Data reduction of XIS \citep{Koya07} and HXD-PIN is 
also the same as \citet{Fuka11a}, following the standard 
{\em Suzaku} data reduction.
We then extracted the spectra of each XIS-FI (0, 2, 3) and XIS-BI (1)
within 4 arcmin of the object, 
and the XIS-FI spectra were coadded; after 2006 Nov., XIS-2
data were not available, and thus we analyzed only XIS-0 and XIS-3 for
the XIS-FI data.
In the line studies, we fitted  XIS-FI and XIS-BI spectra simultaneously with
thr cross normalization set to be free.

For the analysis of Fe-K line variability, we selected AGNs which have
been observed multiply with {\em Suzaku}.  
Furthermore, we gave criteria
that a photon count after background subtraction is more than 2500 in
5--9 keV.  
In addition, we do not include objects with strong Fe-K
absorption lines; NGC 1365, MRK 766, MRK 335, PDS 456, and Fairall 51.
H1821+643 is excluded since the Fe-K$\alpha$ line is not detected for
all observations in the following analysis.  
As a result, 88 observational
data of 25 objects are analyzed for study of Fe-K flux variability.
Table \ref{FeKsample} summarizes the data we analyzed for Fe-K line
variability. 

For the analysis of Ni-K line, we analyzed data whose
observed photon count of XIS-FI after background subtraction is more than
5000 in 5--9 keV.  
Furthermore, we set a criteria for target selection
that the degree of freedom in the fitting of the 7.1--8.0 keV band is
more than 3 (for details, see \S3.2).  
As a result, 143 out of 261
observations remain for studies of Ni-K line.

Since there are only a few pairs of {\em Suzaku} observations with a 
separation of 0.5--1 yr, we analyzed archival {\em XMM-Newton} data of
objects which has been observed by {\em Suzaku} repeatedly
in order to study the Fe-K line variability with a timescale of 0.5--1 yr.
For such objects, we searched for pairs of observations with a
separation around 0.5--1 yr, and found that nine objects have such a pair of
observations.
Table \ref{FeKsampleNewton} summarizes the {\em XMM-Newton} data we
analyzed for Fe-K line variability.  
We performed the standard data reduction with {\tt SAS} 14.0.0.
We extracted the source and background spectra and generated the
response functions with {\tt xmmextractor}.
We fitted MOS-1, MOS-2, and PN spectra simultaneously with 
thr cross normalizations set to be free.

\section{Results}

\subsection{Fe-K line variability}

In order to derive the Fe-K line flux, we fitted the XIS spectra in 5--9
keV.
In addition, we ignored the energy band of XIS-1 where the flux is less
than 0.02 count s$^{-1}$ keV$^{-1}$, to avoid the background uncertainty.
The continuum shape is determined by fitting the 2--10 keV XIS and
15--50 keV PIN spectra by the model including an partially 
absorbed power-law and a reflection component.
Reflection component is modeled by {\tt pexmon} \citep{Nand06} with
the inclination and high energy cut-off fixed to 60$^{\circ}$ and 1000
keV, respectively. 
Partial absorption is modeled by {\tt zphabs} and {\tt pcfabs}.
We determined the continuum parameters, and used them in the fitting
of Fe-K line region (5--9 keV) by fixing the parameters of absorption model of
{\tt zphabs} and {\tt pcfabs}, and
reflection fraction.
Then, we replaced the model {\tt pexmon} by the model {\tt pexrav}
\citep{Magd95}, where only the reflection continuum is considered.
As a result, free parameters of the continuum model are a photon index
and a normalization of the power-law.
In addition, we included five gaussians 
which represent three fluorescence lines of Fe-K$\alpha$ (6.394 keV), 
Fe-K$\beta$ (7.08 keV), 
and Ni-K$\alpha$ (7.47 keV), and two ionized Fe-K$\alpha$ 
lines of He-like (6.7 keV) and H-like (7.0 keV), 
by the model {\tt zgaussian}.
The ionized Fe-K$\alpha$ lines represent ionized emission or
absorption lines observed in many AGNs \citep{Fuka11a}, and we
represent them by a positive or negative normalization of {\tt
zgaussian} model.
 Note that consideration of these two lines in the fitting
introduces some additional errors to the fluorescence Fe-K$\alpha$ 
line intensity.
Energies of these lines other than the 6.4 keV line are fixed to 
the rest-frame energy and shifted by redshift, and
widths of these lines are fixed to 0.
The normalization ratio of the fluorescence Fe-K$\beta$ to the K$\alpha$
line is fixed to 0.125 \citep{Palm03}.
The above modeling is the same as done in \citet{Fuka11a}.
As a result, the fits are almost successful with a reduced $\chi^2$ of $<1.55$.
Two exceptions are  NGC 4151 (2012) and NGC 3783 (2009), 
where a low-energy-tail structure at the left side of the Fe-K$\alpha$ 
line exists and thus their reduced $\chi^2$ is $>$2.
This structure could be reproduced by the relativistic reflection 
model \citep{Keck15,Reyn12}.
Here, we added the broad gaussian but the narrow Fe-K$\alpha$ line flux
does not change beyond statistical errors; we treat only the narrow line
for Fe-K line variability study.
Two observations of NGC 4945 (2010 Jul 9) and MRK 841 (Jul, 2007) give
a somewhat worse fit with a reduced $\chi^2$ of $\sim$2, and thus we do
not treat these data.
We took an average of the line flux of XIS-FI and XIS-BI, weighting them 
at 3 : 1, considering the observed number of photons.
The difference between XIS-FI line flux and the average one is addded to the
statistical error as a systematic error.
In most cases, the difference is at most 2\%, but it is sometimes up tp 5\%.
As a result,
obtained Fe-K$\alpha$ line flux of 88 observations (25 objects) 
is summarized in table \ref{FeKsample}.
When we changed the fitting energy range to 6.0--8.0 keV, 
we obtained consistent Fe-K$\alpha$ flux within errors.
In order to quantify the flux variability of the central engine, we
fitted the PIN spectra in 15--70 keV by power-law and derived the flux
in 20--30 keV.

For {\em XMM-Newton} data, we analyzed the spectra in 5--8 keV by
modeling the absorbed powerlaw plus 5 gaussians {\tt const*(phabs*powerlaw
+ 5*zgaussian)} since the reflection continuum cannot be constrained by
only {\em XMM-Newton} and is not significantly
contributed to the Fe-K band for analyzed data, 
according to the {\em Suzaku} data.
Five {\tt zgaussian} models were treated in the same way as the {\em
Suzaku} data analysis.
All the spectra were well fitted by this modeling with the reduced
$\chi^2<1.1$.
We also checked the pile-up by looking at the plots created with 
the tool {\tt epatplot} and comparing the constant factors of each
instrumental spectra.
Then, the data of the brightest object Cen A suffered significantly 
from the pile-up, and other objects often have a scatter of constant factors
with $\sim$10\% difference between MOS-1 and MOS-2, and often with a larger
difference between MOS and PN.
Therefore, we did not treat Cen A data here.
For other objects,
we took an average of the line flux of MOS-1 and MOS-2, and 
the difference between MOS-1 line flux and the average one is addded to the
statistical error as a systematic error.
The results are summarized in table  \ref{FeKsampleNewton}.

\begin{figure}[lt]
\begin{minipage}{0.5\hsize}
\centerline{\includegraphics*[width=10cm,angle=0]{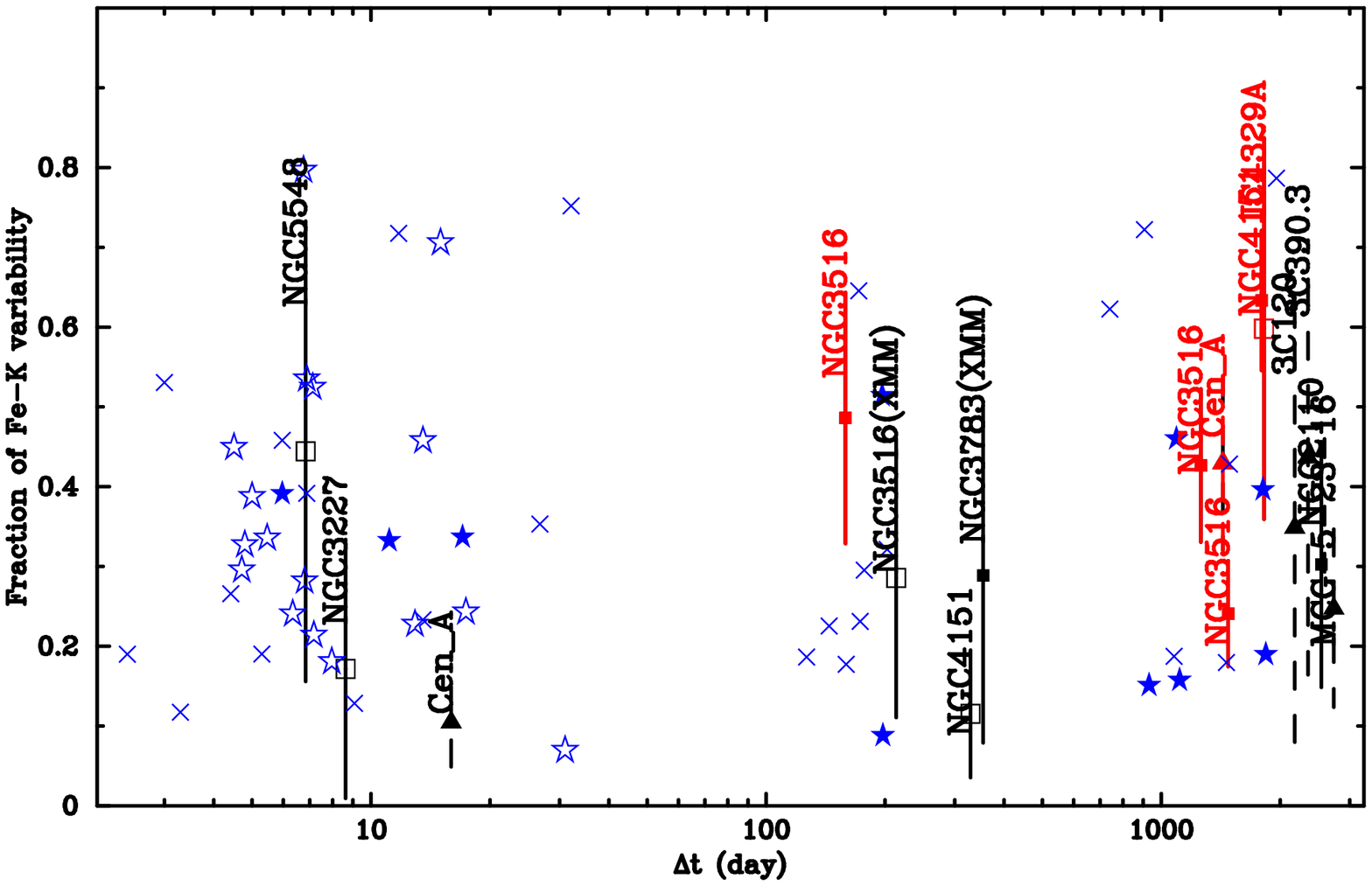}}
\end{minipage}
\begin{minipage}{0.5\hsize}
\centerline{\includegraphics*[width=10cm,angle=0]{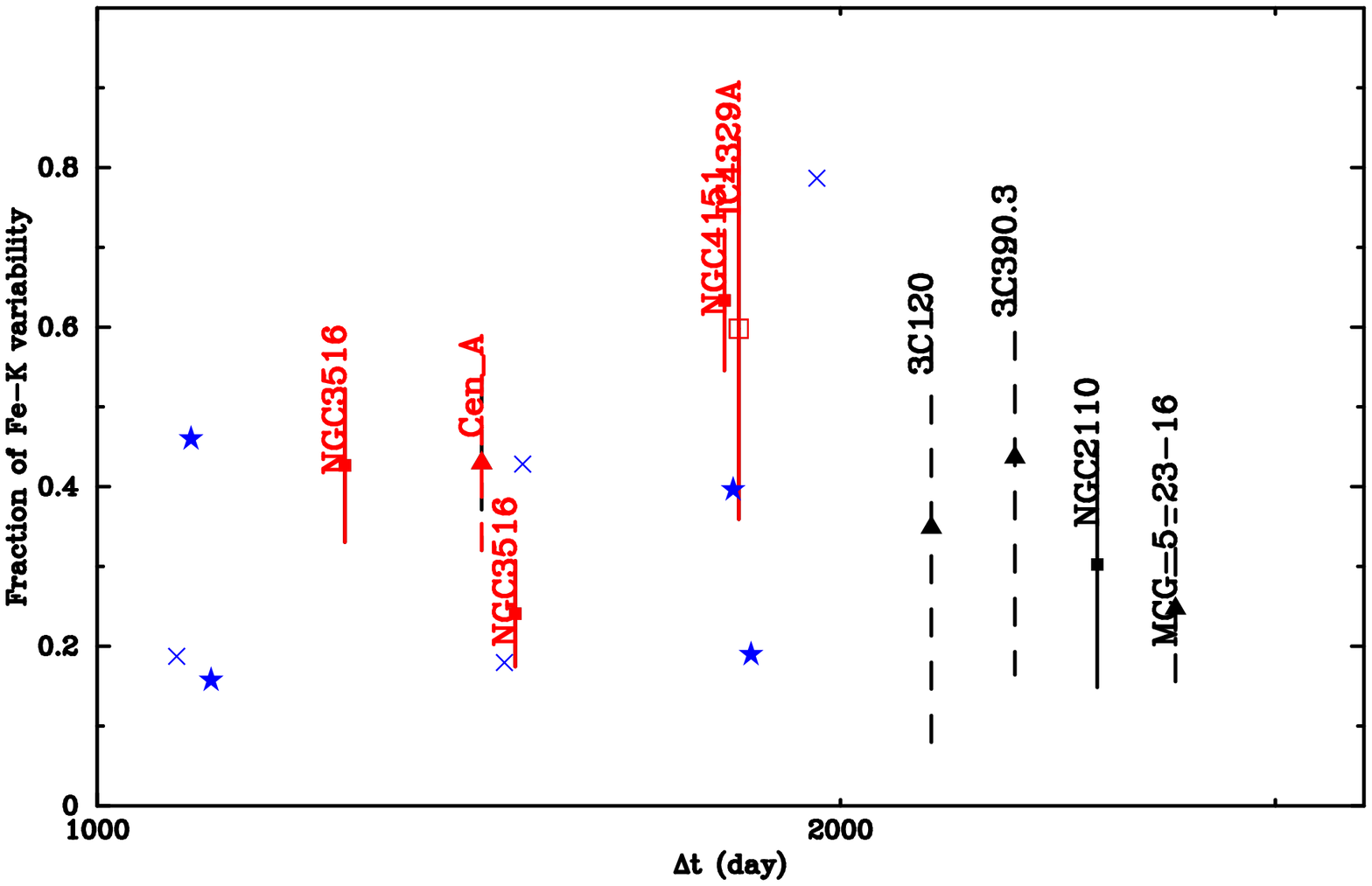}}
\end{minipage}
\vspace*{1.cm}
\caption{Fractional Fe-K variability against the separation time with
 the previous observation. Error bars represent 1$\sigma$.
Red and green data have significance of
 variability with $>2\sigma$ and $(1-2)\sigma$, respectively. 
Squares with solid error bars and stars represent objects whose the 
PIN flux variance
among all observations is $>30$\%, while triangles with
 dashed error bars and crosses represent objects which do not.
Stars and crosses are upper limits for objects with no significant
 Fe-K line variability for PIN flux variance with $>30$\% and $\leq30$\%,
 respectively. 
Filled and open data correspond to objects whose PIN flux variability 
between a pair of observations is $>30$\% or $\leq30$\%, respectively.
Right panel is an enlargement for $\Delta t>1000$ days.
``XMM'' indicates the result derived with the {\em XMM-Newton} data.
}
\label{fek}
\end{figure}

Figure \ref{fek} shows the variability fraction of Fe-K line flux against the
separation time $\Delta t$ for a pair of two sequence observations.
Error bars represent 1 $\sigma$ level, and 
the stars and crosses represent upper limits.
AGNs with significant variability of Fe-K line flux are indicated in 
Table \ref{FeKsample}.
Since the objects have been observed as snapshots, we do not know the
complete history of flux variability of the central engine, 
and thus we cannot consider whether the
objects are expected to show a variability of Fe-K line flux or not.
{\em Suzaku} measured the direct emission flux almost free from the
absorption with PIN, and thus the PIN flux can be used as an indicator
of variability of the central engine.
Therefore, 
we plot the data with different symbols whether the PIN flux variance
among all observations is $>30$\% (square, star) or not (triangle,
cross), 
and whether the PIN flux variability between a pair of observations is
$>30$\% (filled) or not (open).
The latter is a direct evidence of flux variability.
The former ensure that the objects varies significantly even if the
flux does not vary between a corresponding pair of observations.
If the flux variability is not significant, we give an upper limit.
$\Delta t$ is concentrated on $<50$ days and $>1000$ days.
The data with $\Delta t<50$ days are monitoring
observations within one observable window (several months), while
those of $\Delta t>1000$ days are reobservations after several years from the
past observations.
Several pairs of observations show a significant Fe-K time variability
of several tens \% with $>2\sigma$ level for $\Delta t>1000$ days; 
Cen A, IC 4329A, NGC 3516 (2 pairs), and NGC 4151.
On the other hand, some objects show a possible variability with
$(1-2)\sigma$ level
Cen A, NGC 3227, NGC 5548 in $\Delta t<20$ days,
NGC 3516, NGC 3783, NGC 4151 in  $\Delta t=200-350$ days,
and 3C 120, 3C 390.3, MCG-5-23-16, NGC 2110 in $\Delta t>2000$ days.
Fe-K line intensity is high for these objects, but equivalent width is
not so large and exposure time is not so long, and thus
the signal-to-noise ratio is low.
One object, NGC 3516, with a separation of 158 days shows a
significant Fe-K variability of $\sim$50\%.
This timescale is much shorter than those of significant variability
at $\Delta t>1000$ days.
We return to this issue in the discussion.

\begin{table}[!t]
{\scriptsize
\begin{center}
\caption{Summary of {\em Suzaku} multiple observations of AGNs}
\label{FeKsample}
\vspace{0.2cm}
\begin{tabular}{lccp{11cm}} \hline
\hline
Object & $N_{\rm obs}$ & V$^a$ & Date$^b$ (ObsID Fe-flux$^c$ Exposure$^d$ $F_{\rm X}^e$) \\
\hline
1H0323+342 & 2 & & 090726(704034010 3.4$\pm 2.6$ 168.4 6.3) 130301(707015010 3.8$\pm 2.3$ 202.6 3.3) \\ 
1H0419-577 & 2 & & 070725(702041010 5.6$\pm 2.0$ 267.3 8.3) 100116(704064010 6.6$\pm 2.4$ 246.2 6.8) \\ 
3C111 & 4 & & 080822(703034010 19.1$\pm 2.9$ 244.8 11.2) 100902(705040010 24.9$\pm 5.3$ 161.4 24.0) 100909(705040020 22.9$\pm 5.7$ 158.8 29.2) 100914(705040030 25.5$\pm 5.5$ 160.8 25.5) \\ 
3C120 & 3 & V1 & 060223(700001030 22.5$\pm 5.1$ 102.6 20.8) 120209(706042010 30.3$\pm 3.3$ 366.1 21.6) 120214(706042020 29.6$\pm 3.9$ 236.1 21.1) \\ 
3C390.3 & 2 & V1 & 061214(701060010 19.9$\pm 3.4$ 165.7 17.5) 130524(708034010 28.6$\pm 4.2$ 200.7 16.6) \\ 
4C+74.26 & 2 & & 071028(702057010 19.1$\pm 5.4$ 164.2 13.8) 111123(706028010 20.6$\pm 3.9$ 202.8 14.0) \\ 
Cen\_A & 6 & V2 & 050819(100005010 229.5$\pm 13.4$ 164.4 159.3) 090720(704018010 328.1$\pm 21.3$ 125.1 282.0) 090805(704018020 362.5$\pm 19.5$ 102.6 249.7) 090814(704018030 342.9$\pm 18.6$ 111.9 271.6) 130815(708036010 330.3$\pm 45.3$ 21.4 287.6) 140106(708036020 318.8$\pm 43.6$ 14.7 134.6) \\ 
ESO511-G030 & 2 & & 120722(707023020 18.0$\pm 1.9$ 448.1 9.5) 120817(707023030 16.6$\pm 4.6$ 103.8 9.7) \\ 
FAIRALL9 & 2 & & 070607(702043010 27.2$\pm 2.5$ 231.6 10.4) 100519(705063010 25.6$\pm 2.4$ 458.6 13.3) \\ 
IC4329A & 6 & V2 & 070801(702113010 60.2$\pm 13.2$ 34.1 53.9) 070806(702113020 61.1$\pm 13.3$ 39.5 61.7) 070811(702113030 62.8$\pm 13.3$ 37.5 61.1) 070816(702113040 68.2$\pm 13.6$ 33.0 55.6) 070820(702113050 54.9$\pm 10.7$ 33.3 43.0) 120813(707025010 87.7$\pm 7.5$ 235.2 55.3) \\ 
M81 & 2 & & 060508(701022010 6.9$\pm 2.4$ 184.9 4.3) 110915(906004010 7.8$\pm 3.9$ 91.8 3.6) \\ 
MCG-5-23-16 & 3 & V1 & 051207(700002010 80.4$\pm 6.6$ 200.5 49.6) 130601(708021010 100.3$\pm 7.5$ 319.1 51.4) 130605(708021020 99.0$\pm 7.4$ 277.7 47.3) \\ 
MCG-6-30-15 & 3 & & 060109(700007010 23.2$\pm 3.0$ 292.4 16.3) 060123(700007020 23.9$\pm 3.7$ 193.8 17.4) 060127(700007030 22.7$\pm 3.6$ 207.5 18.4) \\ 
MRK509 & 4 & & 060425(701093010 28.0$\pm 10.5$ 36.9 24.0) 061014(701093020 23.0$\pm 7.7$ 59.9 22.9) 061115(701093030 26.8$\pm 11.1$ 27.5 22.3) 061127(701093040 21.3$\pm 8.1$ 46.2 19.1) \\ 
MRK590 & 2 & & 110123(705043010 12.7$\pm 3.1$ 123.2 2.8) 110126(705043020 10.5$\pm 3.4$ 81.8 2.5) \\ 
MRK841 & 4 & & 070122(701084010 13.0$\pm 4.4$ 75.0 7.9) 070723(701084020 12.1$\pm 4.3$ 73.8 8.7) 120105(706029010 13.5$\pm 1.5$ 488.7 5.1) 120118(706029020 13.1$\pm 2.2$ 220.6 6.6) \\ 
NGC2110 & 3 & V1 & 050916(100024010 69.3$\pm 6.3$ 225.5 68.0) 120831(707034010 90.2$\pm 8.6$ 206.5 89.0) 150320(709011010 89.5$\pm 9.7$ 95.1 43.5) \\ 
NGC3227 & 6 & V1 & 081028(703022010 39.1$\pm 4.8$ 117.8 27.7) 081104(703022020 43.2$\pm 5.0$ 78.4 23.2) 081112(703022030 42.8$\pm 5.4$ 75.0 21.4) 081120(703022040 35.4$\pm 4.3$ 73.3 15.3) 081127(703022050 35.9$\pm 5.7$ 65.7 18.9) 081202(703022060 32.8$\pm 5.0$ 66.4 16.3) \\ 
NGC3516 & 8 & V2 & 051012(100031010 56.0$\pm 3.0$ 323.7 22.8) 091028(704062010 42.5$\pm 2.1$ 502.7 10.4) 130409(708006010 24.4$\pm 3.5$ 102.8 3.5) 130427(708006020 24.4$\pm 7.4$ 38.2 10.0) 130512(708006030 32.6$\pm 5.1$ 100.9 11.8) 130523(708006070 35.7$\pm 5.7$ 103.0 15.6) 130529(708006040 37.0$\pm 4.5$ 108.4 11.3) 131104(708006060 19.0$\pm 3.7$ 92.4 2.2) \\ 
NGC3783 & 2 & & 060624(701033010 71.7$\pm 5.2$ 158.4 29.4) 090710(704063010 67.9$\pm 5.5$ 419.8 40.2) \\ 
NGC4051 & 3 & & 051110(700004010 15.7$\pm 2.5$ 255.3 6.4) 081106(703023010 19.3$\pm 2.7$ 373.1 12.0) 081123(703023020 18.8$\pm 3.2$ 111.3 8.5) \\ 
NGC4151 & 4 & V2 & 061218(701034010 169.2$\pm 6.9$ 179.6 52.5) 111117(906006010 276.3$\pm 13.1$ 123.3 171.9) 111218(906006020 276.6$\pm 13.7$ 121.2 179.6) 121111(707024010 244.7$\pm 17.3$ 300.6 165.7) \\ 
NGC4945 & 3 & & 060115(100008030 34.5$\pm 2.5$ 195.2 46.5) 100709(705047020 34.1$\pm 4.2$ 88.3 19.3) 110129(705047050 32.4$\pm 3.7$ 92.2 15.2) \\ 
NGC5506 & 3 & & 060808(701030010 88.0$\pm 9.1$ 96.3 59.3) 060811(701030020 92.1$\pm 8.7$ 115.6 63.6) 070131(701030030 83.3$\pm 9.0$ 77.5 59.8) \\ 
NGC5548 & 7 & V1 & 070618(702042010 19.8$\pm 5.1$ 45.2 7.9) 070624(702042020 19.1$\pm 4.8$ 56.8 10.4) 070708(702042040 20.3$\pm 5.8$ 46.0 12.1) 070715(702042050 22.8$\pm 5.7$ 44.3 13.6) 070722(702042060 19.9$\pm 7.3$ 37.6 16.9) 070729(702042070 26.4$\pm 5.8$ 42.9 14.3) 070805(702042080 14.7$\pm 5.0$ 48.4 11.8) \\ 
\hline
\end{tabular}
\end{center}
$a$: `V1 and V2 denote objects for  which Fe-K line variability is
 detected with 1$\sigma$ or 2$\sigma$ significance  \\
$b$: Observation date with {\tt YYMMDD}. \\
$c$: Fe-K$\alpha$ line flux in unit of $10^{-6}$ photons s$^{-1}$ cm$^{-2}$. \\
$d$: Exposure of XIS-FI (sum of 0, 2, and 3) in unit of kiloseconds. \\
$e$: Hard X-ray flux in 20--30 keV in units of $10^{-12}$ erg cm$^{-2}$
 s$^{-1}$. For NGC 2110, HXD was not operated on the observation
 YYMMDD=150320, we scaled the flux of YYMMDD=120831 by the ratio of
 powerlaw normalization with a photon index of 1.81.
}
\end{table}

\begin{table}[!t]
{\scriptsize
\begin{center}
\caption{Summary of pairs of {\em XMM-Newton} two observations of AGNs}
\label{FeKsampleNewton}
\vspace{0.2cm}
\begin{tabular}{lcp{12cm}} \hline
\hline
Object & V$^a$ & Date$^b$ (ObsID Fe-flux$^c$ Exposure$^d$ $F_{\rm X}^e$) \\
\hline
Cen\_A$^f$ & & 130807(0724060601 104.8$\pm 14.4$  25.9 106.8) 140106(0724060701 164.4$\pm  9.2$  57.1  78.7) \\ 
FAIRALL9 & & 140102(0721110201  20.8$\pm  2.4$ 114.1  11.4) 140509(0741330101  21.6$\pm  1.9$ 336.6  11.6) \\ 
MCG-5-23-16 & & 010513(0112830301  46.1$\pm 10.4$  29.9  35.6) 011201(0112830401  48.6$\pm  6.6$  64.7  34.1) \\ 
MRK509 & & 001025(0130720101  21.7$\pm  3.5$  65.1  15.0) 010420(0130720201  22.4$\pm  4.5$  54.1  18.2) \\ 
NGC3516 & V1 & 010410(0107460601  33.6$\pm  4.3$ 151.3  14.2) 011109(0107460701  43.2$\pm  4.0$ 317.8  16.2) \\ 
NGC3783 & V1 & 001228(0112210101  57.5$\pm  8.9$ 100.1  27.3) 011217(0112210201  44.6$\pm  3.1$ 342.9  17.5) \\ 
NGC4151 & & 060516(0402660101 178.5$\pm  8.1$ 102.7  43.6) 061129(0402660201 174.1$\pm  7.2$  82.3  66.5) \\ 
NGC5506 & & 080727(0554170201  73.4$\pm  4.9$ 225.1  65.5) 090102(0554170101  68.6$\pm  5.5$ 193.0  71.3) \\ 
NGC5548 & & 001224(0109960101  21.2$\pm  3.3$  61.5  11.0) 010709(0089960301  26.4$\pm  4.7$ 216.8  17.2) \\ 
\hline
\end{tabular}
\end{center}
$a$: `V1 and V2 denote objects for  which Fe-K line variability is
 detected with 1$\sigma$ or 2$\sigma$ significance  \\
$b$: Observation date with {\tt YYMMDD}. \\
$c$: Fe-K$\alpha$ line flux in units of $10^{-6}$ photons s$^{-1}$ cm$^{-2}$. \\
$d$: Exposure of sum of MOS-1, MOS-2, and PN in unit of kiloseconds. \\
$e$: Hard X-ray flux in units of $10^{-12}$ erg cm$^{-2}$ s$^{-1}$,
 converted the absorption-corrected 2--10 keV flux to by using the 
powerlaw with a photon index of 1.7. \\
$f$: The data of this object were significantly affected by the pile-up. 
}
\end{table}

\subsection{Ni-K$\alpha$ line intensity}

\begin{figure}[lt]
\begin{minipage}{0.5\hsize}
\centerline{\includegraphics*[width=10cm,angle=0]{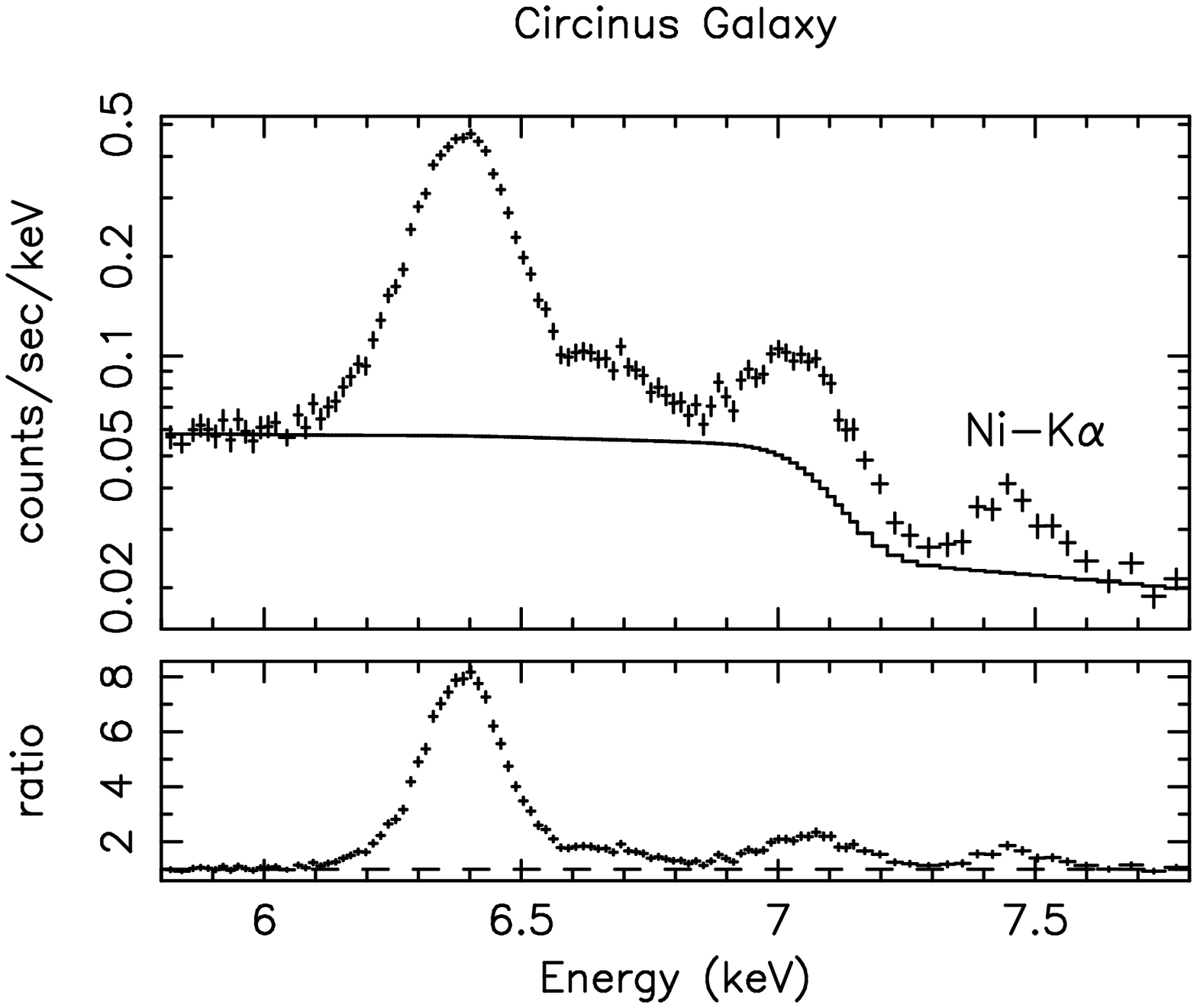}}
\end{minipage}
\begin{minipage}{0.5\hsize}
\centerline{\includegraphics*[width=10cm,angle=0]{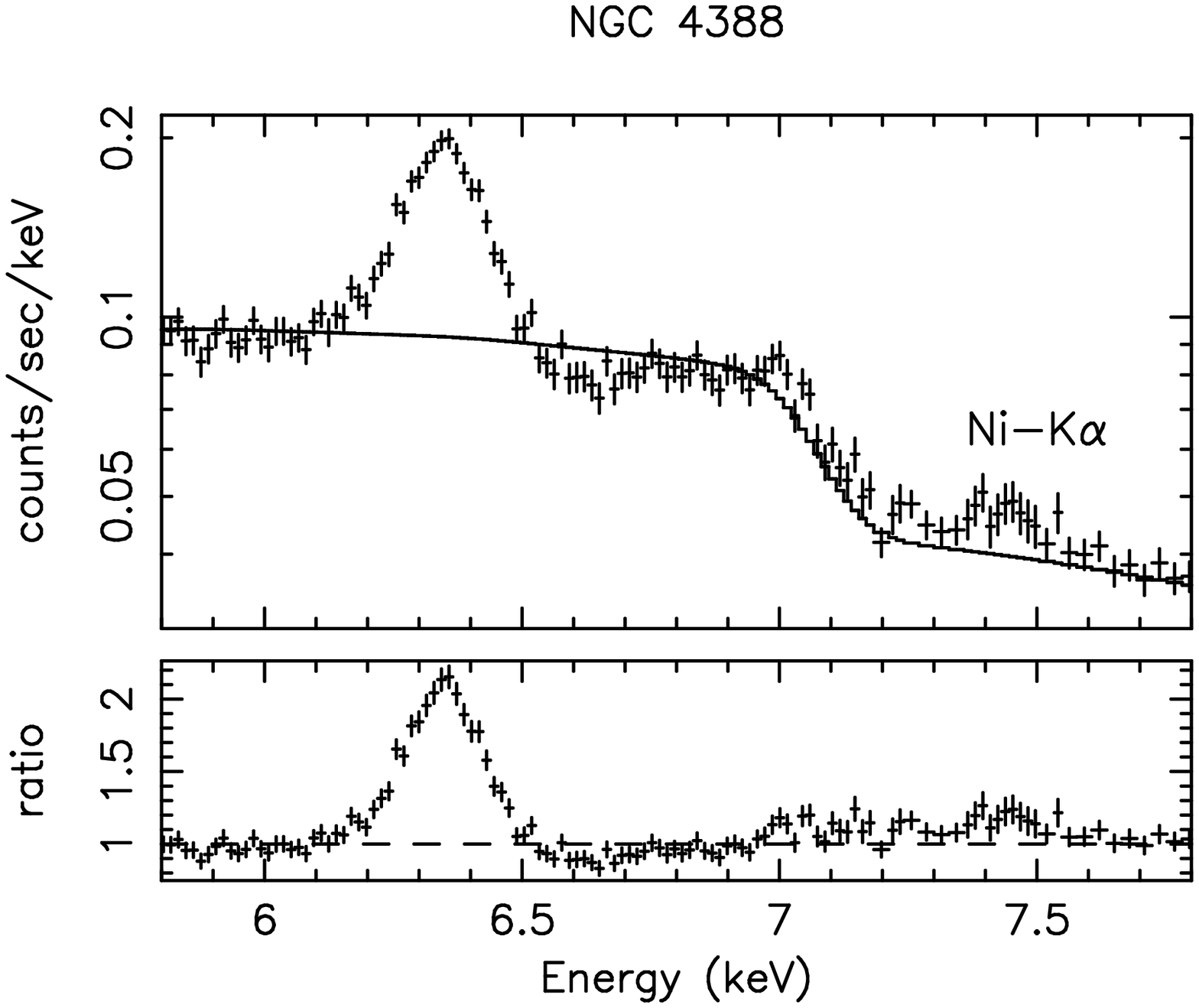}}
\end{minipage}
\vspace*{0.5cm}
\caption{Example of observed Ni-K$\alpha$ line in the XIS spectra;
 Circinus Galaxy (left) and NGC 4388 (right). Solid-line represents the
 baseline continuum model. Bottom panels show a spectral ratio to the
 baseline continuum model.}
\label{nisp}
\end{figure}

\begin{figure}[!t]
\centerline{\includegraphics[width=10cm]{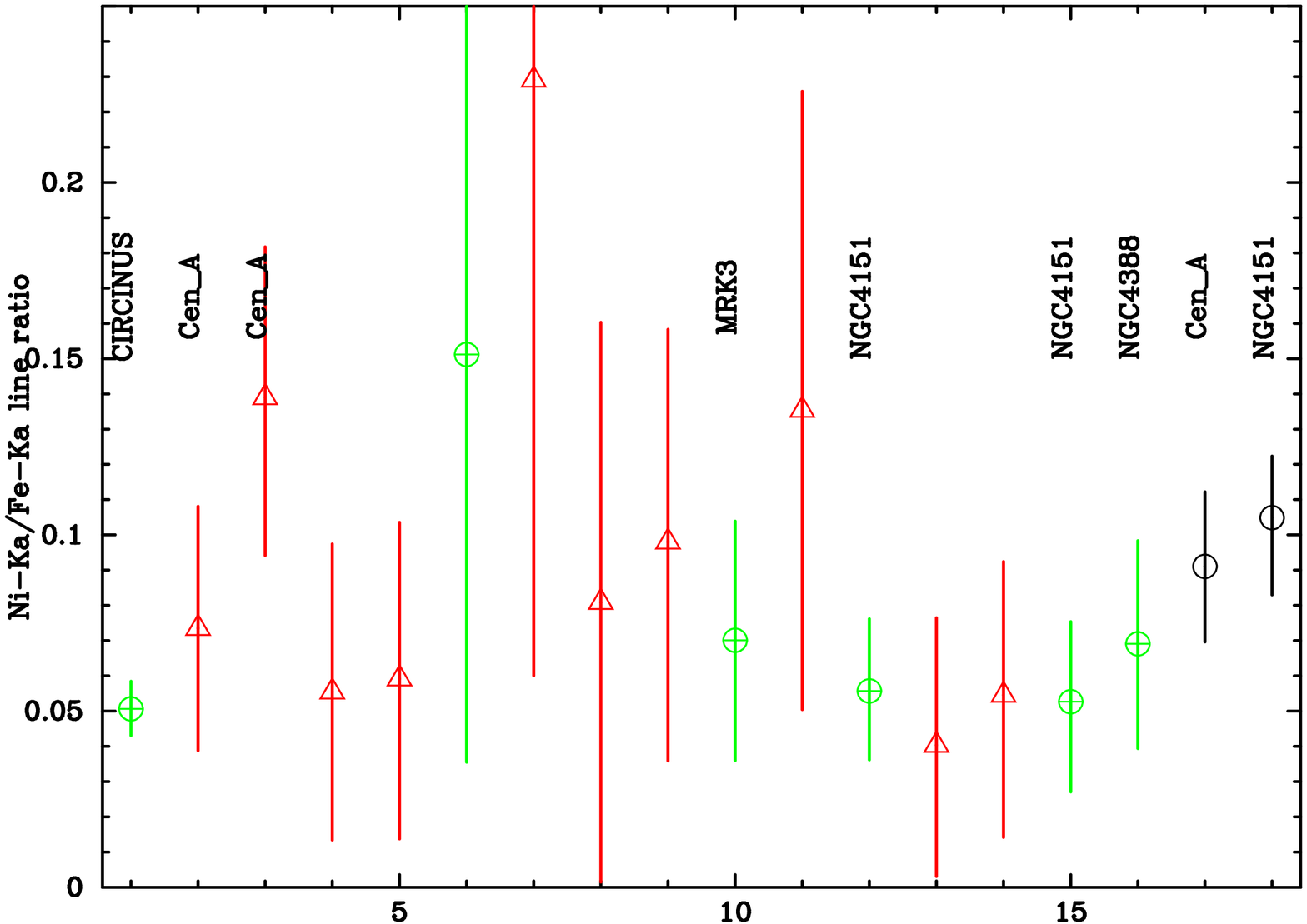}}
\vspace*{0.5cm}
\caption{Ni-K$\alpha$ to Fe-K$\alpha$ line ratio $I_{\rm Ni}/I_{\rm
Fe}$ for objects with a significance of 90\% confidence. Order from left
 to right is the same as that of table \ref{NiKsample}, except for the 2
 black data.
Errors represent 90\% confidence range.
Green and red
 data represent objects with the Fe-K$\alpha$ line intensity of $>10^{-4}$
 photons s$^{-1}$ cm$^{-2}$ or not, respectively. Black data represent
 results of integration of multiple observations. Object name is given
 for objects with $>2\sigma$ significance of Ni-K line detection.}
\label{ni}
\end{figure}

By using the same model as used in the Fe-K line analysis, we obtained the
Ni-K$\alpha$ line flux.
Figure \ref{nisp} shows an example of the observed Ni-K$\alpha$ line.
the Circinus Galaxy,
which has the highest Fe-K line flux \citep{Fuka11b},
shows a clear Ni-K line, while some other AGNs show a marginal Ni-K line as
NGC 4388.
Since a Ni-K$\alpha$ line is very weak (1/20$\sim$1/30 times as intense
as Fe-K$\alpha$ line) , we limited the energy 
band to 7.1--8.0 keV to avoid a situation
in which the continuum level mismatch causes an incorrect 
Ni-K$\alpha$ line flux.
The line parameters other than the Ni-K$\alpha$ line are fixed to those of
Fe-K line analysis.
The spectra are well fitted with a reduced $\chi^2$ of $<2$ 
in all observations.
Although the Ni-K line was not significantly detected for most 
observations due to a weakness of the Ni-K line, 
16 observational data (9 objects) show a 
significance of $>1\sigma$ for Ni-K$\alpha$ line detection.
Figure \ref{ni} plots a line flux ratio of
Ni-K$\alpha$ to Fe-K$\alpha$ line $I_{\rm Ni}/I_{\rm Fe}$ for these 16
observations, and table \ref{NiKsample} summarizes the results.
The Ni-K$\alpha$ line fluxes of Cen A (2005), MRK 3, and the Circinus Galaxy 
are consistent with those in the past results of {\em Suzaku}
observations \citep{Mark07,Awak08,Yang09}.
The average of $I_{\rm Ni}/I_{\rm Fe}$ is 0.066 with a variance of 0.026
for measurements of single observation.
The most accurate measurement was obtained to be 
$I_{\rm Ni}/I_{\rm Fe}=0.0486\pm0.0066$ 
for the Circinus Galaxy, whose equivalent width of the Ni-K$\alpha$ line is
prominently the largest.
Several other AGNs also provide a mild measurement for Cen A, MRK 3,
NGC 4388, and NGC 4151, but the Ni-K line significance is not so high.

When we changed the fitting energy range to 7.1--7.8 keV, 
we obtained consistent Ni-K$\alpha$ flux within errors.
When we modeled the continuum by only absorbed powerlaw ({\tt
phabs*powerlaw} in {\tt XSPEC}) instead of the above model 
(which contains partial covering absorption and reflection), 
Ni-K$\alpha$ flux does not change within errors.
Exceptional cases are the results of Circinus Galaxy and NGC 4151, whose
spectra around the Fe-K edge cannot be well represented by the simple 
absorbed powerlaw model, and thus their Ni-K$\alpha$ flux changes due to
the incompleteness of continuum modeling.

In addition to data analysis of single pointing observations, 
we summed the spectra
of multiple observations for objects which were repeatedly observed,
and analyzed them in the same way as data of single observation.
We also plot the results in figure \ref{ni} (black circle).
Ni-K line intensity ratio of multiple observations is systematically larger
than that of single observation.
For Cen A and NGC 4151, they were observed during the bright state
several times and Ni-K line equivalent width became small.
Therefore, the summation of multiple observations does not necessarily
increase the signal-to-noise ratio and thus does not give a confident result.
Therefore, we do not refer to results of multiple observations.

\begin{center}
\begin{table}[!t]
\caption{Summary of observed Fe-K$\alpha$ and Ni-K$\alpha$ line flux}
\label{NiKsample}
\vspace{0.2cm}
\begin{tabular}{lcccccc} \hline
\hline
Object & Obs$^a$ & Fe-K$\alpha^{b}$ & Fe-EW$^{c}$ & Ni-K$\alpha^{b}$ & Ni-EW$^{c}$ & reduced $\chi^2$ (d.o.f.) \\
\hline
CIRCINUS & 060721 & 337.3$\pm$ 4.9 & 1435$\pm$  21 & $17.1_{-2.5}^{+2.6}$ & $142.4_{-21.3}^{+10.7}$ & 1.52 (23) \\ 
Cen\_A & 050819 & 228.5$\pm$10.7 &   69$\pm$   3 & $16.8_{-7.9}^{+7.9}$ & $7.4_{-3.5}^{+1.7}$ & 1.16 (112) \\ 
Cen\_A & 090720 & 327.1$\pm$17.0 &   63$\pm$   3 & $45.5_{-14.5}^{+13.8}$ & $12.9_{-4.1}^{+5.1}$ & 1.08 (120) \\ 
Cen\_A & 090805 & 361.5$\pm$18.4 &   75$\pm$   4 & $20.0_{-15.2}^{+15.1}$ & $5.9_{-4.5}^{+2.6}$ & 1.05 (116) \\ 
Cen\_A & 090814 & 341.9$\pm$18.4 &   63$\pm$   3 & $20.2_{-15.5}^{+15.1}$ & $5.3_{-4.1}^{+2.3}$ & 1.04 (120) \\ 
Cen\_A & 140106 & 317.8$\pm$41.2 &  103$\pm$  13 & $48.0_{-36.2}^{+37.3}$ & $22.2_{-16.7}^{+23.7}$ & 1.41 (24) \\ 
IC4329A & 070816 & 67.2$\pm$13.1 &   59$\pm$  11 & $15.4_{-10.9}^{+10.9}$ & $19.5_{-13.9}^{+6.1}$ & 1.10 (22) \\ 
MCG+8-11-11 & 070917 & 50.2$\pm$ 4.9 &   68$\pm$   7 & $4.1_{-4.0}^{+4.0}$ & $7.7_{-7.5}^{+0.9}$ & 0.83 (63) \\ 
MCG-5-23-16 & 051207 & 79.4$\pm$ 6.1 &   73$\pm$   6 & $7.8_{-4.9}^{+4.7}$ & $10.2_{-6.4}^{+1.4}$ & 1.27 (77) \\ 
MRK3 & 051022 & 50.2$\pm$ 2.6 &  427$\pm$  22 & $3.5_{-1.7}^{+1.7}$ & $60.2_{-29.1}^{+2.9}$ & 0.71 (9) \\ 
NGC2110 & 150320 & 88.5$\pm$ 8.8 &   94$\pm$   9 & $12.0_{-7.4}^{+7.9}$ & $18.1_{-11.2}^{+4.1}$ & 1.32 (43) \\ 
NGC4151 & 061218 & 168.1$\pm$ 5.3 &  266$\pm$   8 & $9.4_{-3.3}^{+3.4}$ & $22.1_{-7.7}^{+2.2}$ & 0.96 (66) \\ 
NGC4151 & 111117 & 275.3$\pm$12.7 &   92$\pm$   4 & $11.1_{-10.2}^{+9.9}$ & $5.4_{-5.0}^{+1.5}$ & 1.10 (106) \\ 
NGC4151 & 111218 & 275.6$\pm$13.0 &   87$\pm$   4 & $15.0_{-11.1}^{+10.4}$ & $6.9_{-5.1}^{+2.0}$ & 1.31 (107) \\ 
NGC4151 & 121111 & 256.9$\pm$ 7.5 &  102$\pm$   3 & $13.5_{-6.6}^{+5.8}$ & $8.2_{-4.0}^{+1.4}$ & 1.50 (120) \\ 
NGC4388 & 051224 & 77.5$\pm$ 3.3 &  207$\pm$   9 & $5.4_{-2.3}^{+2.3}$ & $23.8_{-10.2}^{+1.5}$ & 0.90 (39) \\ 
\hline
\end{tabular}
\\
$a$: Observation date with {\tt YYMMDD}. \\
$b$: Line flux in unit of $10^{-6}$ photons s$^{-1}$ cm$^{-2}$. \\
$c$: Line Equivalent Width in unit of eV. 
\end{table}
\end{center}

\clearpage

\section{Discussion}

Fe-K line flux of bright AGNs is typically around
$10^{-5}$ photons s$^{-1}$ cm$^{-2}$,
and thus an effective area of $>100$ cm$^{2}$ around 6 keV is needed to
obtain 10\% accuracy with 100 ks observation.
Actually, more effective area is needed to reduce a Poisson noise 
of continuum.
Therefore, {\em Suzaku} XIS and {\em XMM-Newton} provide 
opportunities of Fe-K variability study.
As a result, we obtained the results that some objects show a
significant Fe-K variability of several 10 percents with $\Delta
t>1000$ days.
This is an extensive result of the past reports \citep{Fuka11b,Ursi15,Mari15}
On the other hand, clear evidences of Fe-K variability with $\Delta t<50$ days are not found.
These facts are consistent with the view that a fluorescence narrow Fe-K line
originates at the distant torus.

\begin{figure}[!t]
\centerline{\includegraphics[width=10cm]{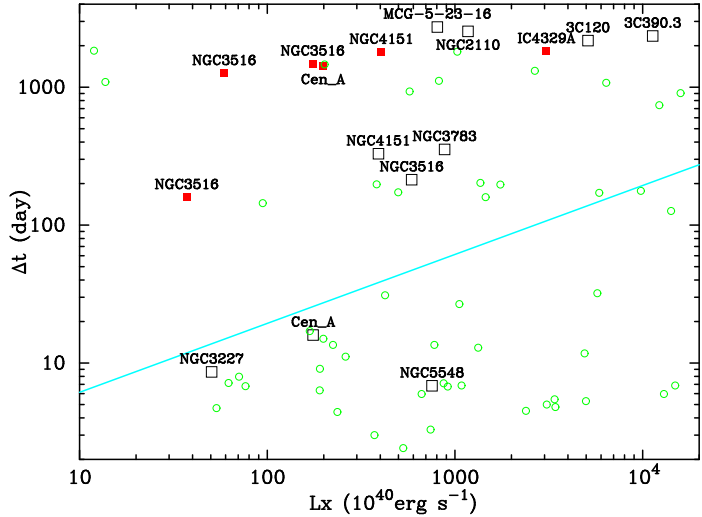}}
\vspace*{0.5cm}
\caption{Observation separation time of a pair of observations against 
X-ray luminosity in 20--30 keV. 
Red squares, black squares, and green circles
 are observations with Fe-K variability of $>2\sigma$, $(1-2)\sigma$,
 $<1\sigma$ level, respectively. A solid line represents an inner radius
 of torus measured by dust echo \citep{Suga06}.}
\label{lxdt}
\end{figure}

Figure \ref{lxdt} plots the separation timescale between a pair of
observations against the X-ray luminosity in 20--30 keV.
For the {\em XMM-Newton} results, we converted the absorption-corrected 
2--10 keV flux to the 20--30 keV luminosity by using the powerlaw with a photon
index of 1.7.
We also plot the inner radius of torus measured by dust echo
\citep{Suga06}; we adjust the scale by NGC 5548 with the V magnitude
of -19.12 \citep{Suga06} and X-ray luminosity (20--30 keV) 
of $6.64\times10^{42}$ erg s$^{-1}$.
Observed timescale of Fe-K line variability is 10--100 times as large
as the inner radius of torus.
Fe-K line variability can appear if the central emission varies
with a large amplitude of $>50$\%, and timescale of such a large
variability is typically around several months.
On the other hand, Fe-K line variability will not be observed when 
the separation timescale between a pair of observations
ish a shorter than the variability timescale of the central emission.
Considering these effects, the separation timescale between a pair of
observations where Fe-K variability is observed is biased to a longer
timescale than a true light-crossing time of Fe-K emitting region.
Therefore, 
observed timescale of Fe-K line variability is larger
than the inner radius of torus.
On the other hand, 
Fe-K line variability will be smeared when the central emission
varies with a shorter timescale than the light-crossing time of
Fe-K emitting region.
The fact that Fe-K line varies with a separation timescale of 1000--2000 days 
between a pair of observations indicates that the Fe-K emitting region
extends to $\sim$1 pc.
Therefore, our results suggest that the Fe-K line emitting region in
some AGNs is
mainly the dust torus within 1 pc of the center, but does not conflict
with some contribution of broad line region to the Fe-K line 
suggested as \citet{Mine15} or \citet{Gand15}.

One special example of significant Fe-K variability is NGC 3516, whose
 Fe-K line intensity varied by $\sim50$\% with $\Delta t=180$ days.
This variability was observed between May and November in 2013, 
when the flux of NGC 3516 was historically low in November.
It is suggested that the inner radius of torus becomes smaller
due to low X-ray luminosity of $(0.5-2.5)\times10^{42}$ erg s$^{-1}$
(15--50 keV) in 2013--2014.
A detail analysis of NGC 3516 will be presented in a forthcoming paper.

CCD energy resolution is around 120 eV at Fe-K line and typical
equivalent widths of Fe-K line for non-Compton-thick Seyfert galaxies 
are 50--200 eV \citep{Fuka11a}.
Therefore, in most cases, the Poisson noise of the continuum significantly 
contributes to the line flux accuracy.
{\em ASTRO-H} SXS has an unprecedentedly good energy resolution of 4--6 eV
\citep{Taka14}, enabling us to measure the Fe-K line flux more
accurately without contribution of continuum.
Also, {\em ASTRO-H} SXS could measure the Fe-K line width with good accuracy, 
and time variability study could track the width variation and thus the
 change of inner torus radii.

{\em Suzaku} XIS observations also enabled us to measure Ni-K$\alpha$ line
flux of AGNs systematically for the first time, due to its
stable and low background.
$I_{\rm Ni}/I_{\rm Fe}$ is 0.03--0.07 for objects with Ni line
detection.
These are almost consistent with the past observations of {\em XMM-Newton} and
{\em Suzaku}, but smaller than 0.13$\pm$0.03 of {\em XMM-Newton} 
measurements for NGC 1068 \citep{Matt04}. 
Since Ni-K line is very weak, it could be affected by background
subtraction or complex absorption features by highly ionized material.
The former case would not be worried about in our results, since most
objects with Ni-K line detection are apparently bright.
Here, 
we discuss the Ni-K$\alpha$ intensity ratio to the Fe-K$\alpha$, by
considering the latter case.

The predicted Ni-K line intensity in the reflection spectrum of AGNs has been
reported by \citet{Yaqo11} in detail, based on their {\tt MYTorus}
model.
They assumed that the Ni-to-Fe abundance ratio is the same as that of 
\citet{Ande89} (Fe and Ni abundances against Hydrogen as $4.68\times10^{-5}$ 
and $1.78\times10^{-6}$, respectively), 
fluorescence yield of K lines is 0.347 and 0.414 for Fe and Ni,
respectively, the K$\alpha$ branching fraction is 0.8811 for both of Fe
and Ni.
Comparing their Ni-K line prediction (figure 1 in their paper), together
with their Fe-K line prediction (figure 2 in \citet{Yaqo10}), 
the Ni to Fe K$\alpha$ line ratio is
0.030--0.033 and 0.032--0.038 for various inclination angles 
for the torus column
density of $10^{24}$ cm$^{-2}$ and $10^{24.5}$ cm$^{-2}$, respectively.
We also predicted the observed Ni to Fe K$\alpha$ line ratio by
simulating the X-ray reflection spectrum with the model {\tt
pexmon} \citep{Nand06}, in
which the Ni to Fe K$\alpha$ line ratio is assumed to be 0.050.
We set an inclination angle as $\cos\theta =0.15$, a metal abundance of 1
solar, and an incident powerlaw photon index of 2.
We convolved the simulated spectra with the XIS response, and 
fitted them in the same way as observed data, 
and obtained a line ratio of 0.051.
This is larger than that based on \citet{Yaqo11},
possibly due to the assumption of larger Ni
abundance and/or Ni fluorescence yield for the database used in {\tt pexmon}.

As an independent check, 
we performed the Monte-Carlo simulation of X-ray reflection by torus 
using the Geant 4 package, as performed by \citet{Odak11} or
\citet{Liu14}, using the framework {\tt MONACO} \citep{Odak11}.
Detailed description on this Monte-Carlo simulation will be presented in
\citet{Furu15}.
The geometry is almost the same as that of \citet{Murp09}.
We assume Fe and Ni abundances, according to \citet{Ande89}.
Fluorescence yield (K$\alpha$ branching fraction) of K lines is 0.3401
(0.8830) and 0.4060 (0.8825) for Fe and Ni,
respectively \citep{Furu15}.
The absorption column density of the torus is assumed to be 
$10^{24.5}$ cm$^{-2}$ toward the equatorial direction.
The incident X-rays follow a power-law shape with a photon index of 2
in the range of 2--300 keV.
The reflection spectrum is accumulated with a viewing angle of
$\theta$ as $0.1\leq\cos\theta\leq0.2$, 
convolved with the XIS response function, and fitted in
the same way as observed data.
As a result, a Fe-K$\alpha$ (6.4 keV) to Ni-K$\alpha$ line intensity
ratio becomes 0.035, almost consistent with that based on \citet{Yaqo11}.
When we changed the assumed Ni abundance to 0.5 solar or 2.0 solar, the
Fe-K$\alpha$ to Ni-K$\alpha$ line intensity ratio varies almost
proportionally to the assumed Ni abundance.
Since the assumed abundance and fluorescence yield of Ni in ours and 
\citet{Yaqo11} is explicitly shown and typically
used, we take the Ni to Fe K$\alpha$ line ratio of 0.035$\pm$0.003 for
the solar abundance.
In this case, the average of observed Ni to Fe K$\alpha$ line flux
ratios of $0.062\pm0.026$ indicates
that the Ni to Fe abundance ratio is 1.9$\pm$0.8 solar, and it is
marginally consistent with but somewhat higher than the solar
abundance ratio.
The Ni to Fe abundance ratio was reported to be larger by a factor of
1.5--2.0 than that of \citet{Ande89} for the
Circinus galaxy \citep{Mole03}, and thus consistent with ours.
This suggests possible evidence of enhanced Ni to Fe abundance ratio
for the torus material of AGNs.
Note that {\em XMM-Newton} results of NGC 1068 \citep{Mole03} lead to 
the extremely large Ni to Fe abundance ratio of 3.7$\pm$0.9 solar.

However, as described in \S3.2, we must pay attention to the continuum
modeling in measurements of a weak Ni-K$\alpha$ line.
Possible systematics are caused by blue-shifted Fe-K absorption lines.
Typical weak blue-shifted Fe-K absorption lines have an equivalent width
of 10--30 eV \citep{Tomb10}.
If such absorption lines exist in the CCD spectra 
but are not modeled in the narrow-band
spectral fitting around the Ni-K line, the continuum would be
underestimated and thus 
the Ni-K line equivalent width would be affected by $\sim$10 eV.
Equivalent widths of observed Ni-K$\alpha$ lines are typically 5--20 eV,
and thus could be overestimated by several percents.
This indicates that the measurement with higher spectral resolution
is important to determine a Ni-K line flux unambiguously.
Therefore, {\em ASTRO-H} SXS \citep{Taka14} observations of AGNs are promising 
for Ni-K line probing.

The enhanced abundance of Ni has recently been 
reported for a supernova remnant 3C 397 whose progenitor 
is considered to be a SN Ia
\citep{Yama15}, who suggested a production of $^{58}$Ni in the 
single-degenerated binary.
The enhanced Ni to Fe ratio has also been reported for three
core-collapse supernovae \citep{Jerk15a} and gives constraint on the
supernova explosion mechanism and structure of parent stars.
\citet{Jerk15b} suggested that stars with a zero-age main sequence star 
(ZAMS) mass of 13--15 $M_{\odot}$ could produce an enhanced Ni to Fe abundance.
If the enhanced Ni to Fe abundance of the torus is significant, more
single-degenerated binary and more middle-mass stars are suggested to 
contribute to the
enrichment of metals at the center of AGN-hosting galaxies.
These issues are related with star formation phenomena, which are
often associated with AGN activity.
Therefore, information of Ni to Fe abundance ratio in other galaxies
and AGNs is important to understand star formation history of
galaxies and AGNs.

The authors also wish to thank all members of the {\em Suzaku} Science Working
Group, for their contributions to the instrument preparation, spacecraft
operation, hardware/software development, and in-orbit calibration. 
Y.F. is supported by JSPS KAKENHI Grant Number 2424401400.
H.N. is supported by the Grant-in-Aid for Young Scientists (B) 
(26800095) from the Japan Society for the Promotion of Science 
(JSPS), and the Special Postdoctoral Researchers Program in RIKEN.

\end{document}